\documentclass[doublecol]{epl2} 

\title{Diode for Bose-Einstein condensates}
\shorttitle{Title} 

\author{Jonas Larson}

\institute{Department of Physics, Stockholm University, 
106 91 Stockholm, Sweden}


\abstract{Given a quantum state at some instant of time $t$, the underlying system Hamiltonian can not only predict how the state will evolve, but also the history of the state prior to $t$. Thereby, in order to have a directed motion, like in a diode, some sort of irreversibility must be considered. For the atom diode, this has been achieved by spontaneous decay of excited atomic levels. For an atomic condensate, it is clear, however, that such decay will cause both heating and decoherence of the condensate. To overcome this complication we introduce a different setup where the dissipation does not act directly on the atoms. The excited atoms are stimulatedly driven back to the ground state by exciting a cavity mode, which in return decays to the vacuum via photon losses. The efficiency of the method utilizing experimental parameters is shown to be almost perfect within large parameter regimes. }

\begin{document}

\maketitle

\section{Introduction}
With the advances in atomic cooling~\cite{cool} and in the construction of atomic chips~\cite{chip}, it is nowadays possible to coherently guide and manipulate atomic matter waves~\cite{atomopt,control}. Retaining the inherent coherence of the matter waves is a necessary condition for successful applications. This becomes maybe especially important in schemes of quantum information processing with neutral atoms~\cite{qinfo} and for quantum memories~\cite{qmemory}. The building blocks in these atom optics models are formed by light-matter interaction in which the light fields constitute effective matter wave potentials. By carefully designing the light-matter interaction one can achieve analogs of for example lenses~\cite{lens}, transistors and amplifiers~\cite{trans}, and atom diodes~\cite{diode,diode2,diode3}. 

In this paper we consider an atom diode -- a device that allows atoms in their ground state to pass in one direction but not in the opposite direction -- where decoherence due to irreversible losses are greatly suppressed. One idea of using atomic Bose-Einstein condensates (BEC) in atom optics setups is the fact that effective time-scales (often scaling inversely with the number of atoms) can be considerably decreased, and moreover, the many atomic internal degrees of freedom of the BEC permitting storing of a large amount of quantum information. However, condensation relies on coherence and it is thereby desirable to minimize the amount of decoherence in the atom diode. Our idea is based on the one of ref.~\cite{diode} (see also~\cite{diode4}), but we remove atomic spontaneous emission and replace it by Raman coupling the atoms to a high-$Q$ optical cavity. Spontaneous emission, as utilized in ref.~\cite{diode}, will induce both heating of the condensate as well as loss of inherent coherence, while in our scheme the atoms follows a dark state during the Raman process and the atomic motion is, within the adiabatic regime, constant. In addition, the irreversible process occurs between the cavity field and its surrounding photon reservoir, and consequently there are no direct decoherence effects on the condensate. Employing experimentally relevant parameters, we find that the diode efficiency is close to perfect within a mean-field treatment. Non-adiabatic contributions, arising either due to high atomic velocities or strong non-linearities, result in atom-field entanglement. These correlations decay thanks to the reservoir-induced collapse of the cavity field, which therefore improves the overall efficiency. However, tunneling resonances may cause non-zero transmission of back scattering atoms. This, on the other hand, can be controlled, and consequently suppressed, by adjusting  the external lasers.

\section{Model system}
\begin{figure}
\includegraphics[width=6cm]{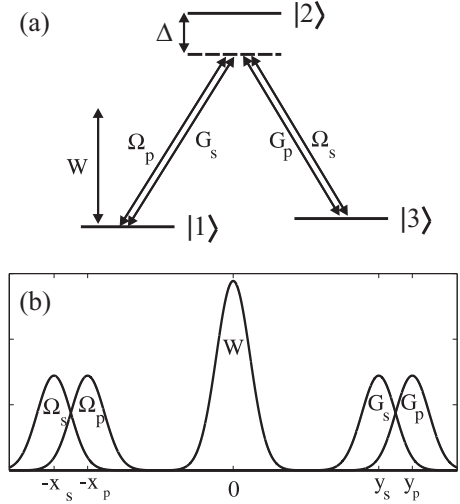}
\caption{ Atomic level configuration (a) and positions of the lasers along the waveguide (b). The two pulses $\Omega_p$ and $G_s$ couple the ground state $|1\rangle$ to the excited state $|2\rangle$, while the remaining two pulses $\Omega_p$ and $G_s$ couple instead the atomic states $|2\rangle$ and $|3\rangle$. The detuning $\Delta$ prevents population of the excited state for atoms propagating in the negative direction. The Stark-shift pulse $W$, acting solely on the atomic $|1\rangle$-state, has twice the amplitude of the Raman pulses, while all the widths are the same. }
\label{fig1}
\end{figure}

We consider a BEC of $^{87}$Rb-atoms propagating along the positive direction in a waveguide aligned in the $x$-direction, i.e. the dynamics in the transverse directions can be frozen out. However, the transverse confinement is still not large enough to cause neither {\it confinement-induced resonances}~\cite{conf} nor collapse of the condensate~\cite{pethick}. Two atomic ground state Zeeman levels $|1\rangle$ and $|3\rangle$ (typically taken from the $5S_{1/2}$ $F=1$ Zeeman manifold) are Raman-coupled to an excited state $|2\rangle$. Two spatially separated classical lasers drive the $2\leftrightarrow3$ transition and another classical laser the $1\leftrightarrow2$ transition. In addition, the atomic states $|1\rangle$ and $|2\rangle$ are also coupled via a mode of an optical Fabry-Perot resonator. The atomic-laser couplings are schematically presented the level diagram of fig.~\ref{fig1} (a). All coupling fields propagate in the transverse direction and thereby possess Gaussian profiles in the longitudinal $x$-direction. Furthermore, they are all detuned from their respective atomic transitions by $\Delta$. Finally, the state $|1\rangle$ is Stark-shifted by yet another classical laser. The positions of lasers along the waveguide and their relative strengths are displayed in fig.~\ref{fig1} (b). Thus, the condensate traverses the pulses in the order; a Stokes ($s$) pulse coupling states 2 and 3, a pump ($p$) pulse coupling 1 and 2, the Stark pulse acting on state 1, followed by another Stokes pulse formed by the cavity and coupling states 1 and 2, and finally a pump pulse between 2 and 3. The amplitude of the Stark-shift pulse $W$ has been taken twice as big compared to those of the Raman Stokes and pump pulses $\Omega_{s,p}$ and $G_{s,p}$,
\begin{equation}\label{pulses}
\begin{array}{l}
W(x)=2\hbar\Omega_0e^{-\frac{x^2}{2w^2}},\\ \\
\Omega_\alpha(x)=\hbar\Omega_0e^{-\frac{(x-x_\alpha)^2}{2w^2}},\hspace{1cm}\alpha=p,\,s,\\ \\
G_\beta(x)=\hbar\Omega_0e^{-\frac{(x-y_\beta)^2}{2w^2}},\hspace{1cm}\beta=p,\,s.
\end{array}
\end{equation}
The pulse positions $x_s$, $x_p$ and $y_s$, $y_p$ are chosen such that population is adiabatically swapped between the two Zeeman ground state levels $|1\rangle$ and $|3\rangle$ via the {\it stimulated Raman adiabatic passage} (STIRAP) process~\cite{stirap}. To achieve a robust population transfer, the two Raman pulses should occur in a counterintuitive way and overlap~\cite{stig}. The advantages of an adiabatic scheme, provided the evolution is adiabatic, is that it is insensitive to parameter strengths as well as effective interaction times. Yet another advantage using a STIRAP scheme is that the evolution follows a dark state $\psi_d(x)$ such that the condensate velocity is ideally not affected by the couplings. As long as we consider ultracold atoms, the center Stark-shift pulse hinders atoms in state $|1\rangle$ to pass. However, atoms entering from the left in state $|1\rangle$ will first be STIRAP transfered into $|3\rangle$-state atoms before arriving at the center pulse, and hence pass the barrier unaffected. The second STIRAP sequence reverses the population swapping, but this time the Stokes pulse consists of a high-$Q$ optical cavity mode such that each atom passing it leaves behind a photon. On the time-scale of the atomic motion, the photons rapidly escape the cavity leaving it in vacuum. Therefore, $|1\rangle$ atoms coming from the right cannot be transfered into the $|3\rangle$ states since the cavity is empty from photons. Atoms arriving from the right in the $|3\rangle$ state, on the other hand, couple to the excited $|2\rangle$ atomic state via the Pump pulse $G_p(x)$. However, this transition is highly detuned due to the non-zero $\Delta$ and hence the population transfer greatly suppressed. The atoms in this state, nevertheless, experience an effective repulsive Stark shift potential upon which they are reflected. Consequently, $|1\rangle$ atoms entering from the left are reflected by the $W(x)$ Stark shift pulse, while $|3\rangle$ atoms are reflected by the detuned $G_p(x)$ pulse. The amount of thermally excited atoms in the $|2\rangle$ state is assumed vanishingly small. 

For the $F=1$ manifold of $^{87}$Rb atoms, the $s$-wave spin-dependent interactions is a factor 200 smaller than the spin-independent interactions~\cite{int}. In the present situation, the effect of interactions plays a minor role and we can safely neglect the spin-dependent interactions. Denoting the $s$-wave scattering length as $a_s$, the atomic mass as $m$, and the transverse condensate widths $\Delta_t$, the effective atom-atom interaction reads $U=2\hbar^2a_s/m\Delta_t^2$~\cite{pethick}. Within a mean-field approach, the Gross-Pitaevskii equation in the rotating wave approximation, 
\begin{equation}
\begin{array}{lll}
\displaystyle{i\frac{\partial}{\partial t}\Psi(x,t)} & = & \displaystyle{\Bigg[-\frac{\hbar^2}{2m}\frac{d^2}{dx^2}+W(x)\hat{\sigma}_{11}+\Delta\hat{\sigma}_{33}},\\ \\
& & \displaystyle{+(\Omega_s(x)+G_p(x))\left(\hat{\sigma}_{23}+\hat{\sigma}_{32}\right)}\\ \\
& & \displaystyle{+\left(\Omega_p(x)+G_s(x)\hat{a}^\dagger\right)\hat{\sigma}_{21}}\\ \\
& & \displaystyle{+\left(\Omega_p(x)+G_s(x)\hat{a}\right)\hat{\sigma}_{12}}\\ \\
& & \displaystyle{+UN|\Psi(x,t)|^2\Bigg]\Psi(x,t)}\\ \\
& = & \hat{H}_{GP}\Psi(x,t),
\end{array}
\end{equation}       
render the time-evolution of the order parameter. Here, the spin operators $\hat{\sigma}_{ij}=|i\rangle\langle j|$, $N$ is the total number of atoms, $\hat{a}$ and $\hat{a}^\dagger$ represent the photon annihilation and creation operators of the cavity mode, and the spinor order parameter
\begin{equation}
\Psi(x,t)=\left[\begin{array}{c}
\psi_{10}(x,t)\\
\psi_{11}(x,t)\\
\psi_{20}(x,t)\\
\psi_{21}(x,t)\\
\psi_{30}(x,t)\\
\psi_{31}(x,t)
\end{array}\right],
\end{equation}
where $|\psi_{ij}(x,t)|^2$ characterizes the atomic density of level $|i\rangle$ and with an empty $j=0$, or populated cavity $j=1$. We have normalized the order parameter to unity, and thereby the appearance of the factor $N$ in the interaction term. Note that the coupling schemes of the Stokes and the pump pulses have been swapped between the two Raman processes in order to reverse the population transfer among the Zeeman levels $|1\rangle$ and $|3\rangle$.

We have treated the condensate at a mean-field level, while the cavity field is at a quantum level. We could replace the operators $\hat{a}$ and $\hat{a}^\dagger$ by the coherent amplitudes $\alpha$ and $\alpha^*$ respectively in the equation above. However, when treating cavity losses within a master equation approach we need to keep the quantum nature of the cavity field~\footnote{Solving the effective master equation takes into account for reservoir-induced quantum noise, which typically is absent in a full mean-field approach.}. We take $\hat{a}$ and $\hat{a}^\dagger$ as the regular boson operators, but it is understood that while the condensate traverse the cavity field its true amplitude is proportional to the atom number $N$. In our model, the cavity field amplitude is independent of the atomic number $N$ and instead $|\alpha|^2=1$. The effective model that we consider for the cavity field should not be a sever shorting since the cavity field decay is independent on the field amplitude, namely $\alpha(t)=\alpha(0)e^{-\kappa t}$ where $\kappa$ is the decay rate. Moreover, The crucial aspect is not in the actual cavity decay as long as it is fast compared to the time-scale set by the condensate motion.

As pointed out in the introduction, for a diode we should have a directed motion of the condensate, i.e. if we reverse time we should not be able to propagate to negative $x$'s provided the condensate is located at large positive $x$'s. Despite rendering unitary evolution, the Gross-Pitaevskii equation is inherently non-linear and time-reversal may cause hysteresis effects. This appears only for very high atomic densities or strong atom-atom interactions, which does not apply to our model. The irreversible dynamics arises due to cavity losses, which is here modeled by a master equation. For the density operator initially in the state $\rho=\Psi^\dagger(x,0)\Psi(x,0)$, and for coupling to a zero temperature photon bath, the equations of motion takes the form~\cite{walls}
\begin{equation}\label{master}
\displaystyle{\frac{\partial}{\partial t}\rho}=\frac{i}{\hbar}[\rho,\hat{H}_{GP}] \displaystyle{+\frac{\kappa}{2}\left(2\hat{a}\rho\hat{a}^\dagger-\hat{a}^\dagger\hat{a}\rho-\rho\hat{a}^\dagger\hat{a}\right)}.
\end{equation}
Losses of the atomic Zeeman levels have been neglected as the excited $|2\rangle$-state is negligible populated within the adiabatic regime that we consider.

\begin{figure}
\includegraphics[width=8cm]{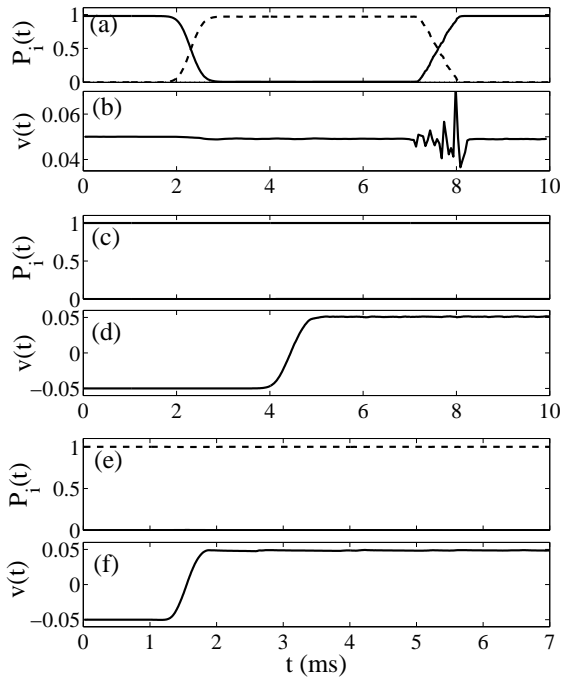}
\caption{ The plots (a), (c), and (e) show the populations $P_i(t)$ of the different Zeeman-levels $|1\rangle$ (solid curve), $|2\rangle$ (dotted curve), and $|3\rangle$ (dashed curve). The condensate velocity $v(t)$ is depicted in (b), (d), and (f). In the first two plots we give the result for atoms propagating from left to right. In this case, the first STIRAP-process results in almost perfect population transfer, indicating the adiabatic following of the dark state. The velocity fluctuations in the second STIRAP-process is due to transitions out of the dark state induced by the cavity decay. The results of atoms initially in state $|1\rangle$ propagating in the negative direction (from right to left) are shown in (c) and (d). Here it is clear how the atoms remain in the same Zeeman-level and how they are reflected by the $W(x)$ pulse. Finally, in (e) and (f) we present the same but for atoms initially in the $|3\rangle$ state. This time the atoms are reflected already at the $G_p(x)$ pulse. The parameters are given in the text.  }
\label{fig2}
\end{figure}

\section{Results}
The numerics will be for a $^{87}$Rb condensate as for the BEC-cavity experiments of Ref.~\cite{esslinger}. More precisely, we take a condensate of $N=100\,000$ atoms, the transverse widths $\Delta_t=3$ $\mu$m, scattering length $a_s=5.77$ $n$m, atom-field couplings $g=\Omega_0=2\pi\times10.9$ MHz, detuning $\Delta=2\pi\times55$ MHz, cavity mode waist $w=15$ $\mu$m, and cavity loss rate $\kappa=2\pi\times1.3$ MHz. Note that via eq.~(\ref{pulses}), the atom-field coupling and the mode waist set the amplitudes and waists for the classical lasers as well. The pulse locations are $x_p=-y_s=-130$ $\mu$m and $x_s=-y_p=-160$ $\mu$m. As an initial state for a condensate propagating in the positive direction we take a Gaussian located at $x_0=-260$ $\mu$m, with a longitudinal width $\Delta_l=10$ $\mu$m, an initial velocity in the positive $x$-direction of $v_0=5$ cm/s~\cite{velo}, and only the $|1\rangle$-component populated
\begin{equation}
\Psi(x,0)=\frac{1}{\sqrt[4]{\pi\Delta_l^2}}e^{imv_0x/\hbar}e^{-\frac{(x-x_0)^2}{2\Delta_l^2}}\left[\begin{array}{c}
1\\
0\\
0
\end{array}\right].
\end{equation}
The actual shape of the condensate has little influence on the protocol, i.e. we could equally well have chosen some sort of Thomas-Fermi profile. The cavity mode is initially in vacuum. For verifying the diode efficiency we also consider condensates propagating in the negative direction. Then the initial state is the same but with $x_0\rightarrow-x_0$ and $v_0\rightarrow-v_0$. we also check for negatively propagating atoms initially in the $|3\rangle$ state and with the same motional wave packet.

We use a Monte-Carlo method~\cite{mc} combined with the split-operator method~\cite{split} to solve the master equation (\ref{master}). The initial state is propagated with the non-Hermitian Hamiltonian $\hat{H}=\hat{H}_{GP}-\frac{i\kappa}{2}\hat{a}^\dagger\hat{a}$. This propagation is performed on some time-grid with steps $dt$. For this we utilize the split-operator method, and chose $dt$ small enough such that convergence is guaranteed. At each time-step we randomly simulate a quantum-jump describing loss of a photon. One such simulation render a single quantum-trajectory. Repeating this procedure and average over all trajectories results in the desired solution. 

We focus on the population of Zeeman levels
\begin{equation}\label{pop}
P_i(t)=\mathrm{Tr}\Big[\hat{\sigma}_{ii}\rho(t)\Big],
\end{equation}
the velocity 
\begin{equation}\label{velo}
v=\frac{d\bar{x}(t)}{dt},
\end{equation}
where $\bar{x}(t)=\mathrm{Tr}\Big[x\rho(t)\Big]$, and the reduced densities
\begin{equation}
\rho_i(t)=\mathrm{Tr}_{j\neq i}\Big[\rho(t)\Big].
\end{equation}
In this last equation, trace is over the cavity field and the remaining two Zeeman levels' degrees of freedom, i.e. $\rho_i(t)$ represent the atomic density of Zeeman level $|i\rangle$. Since the atomic velocity should ideally not couple to any other degrees-of-freedom, the velocity is an indirect measure of adiabaticity. More precisely, the initial state populates a dark state with the corresponding adiabatic potential~\cite{jonas} being strictly zero, and as long as the dynamics is adiabatic the velocity should stay constant. On the other hand, if any of the bright states become populated due to non-adiabaticity it will render a velocity change since their corresponding adiabatic potentials are non-zero and thereby generates an effective force. The bright states normally contain as well population of the Zeeman state $|1\rangle$ which will be reflected upon the center pulse $W(x)$. In addition to the above quantities, we will also calculate the total transmission probability $T$ in order to have a second verification of the directed motion of the condensate.  

Figure~\ref{fig2} presents various numerical results of the populations and the velocities, obtained using the above mentioned parameters and averaging over 500 Monte Carlo trajectories~\footnote{For backward propagation, the cavity stays in vacuum and no quantum jumps occur. Consequently a single trajectory captures the evolution in these cases.}. The first two plots, (a) and (b), give the results for atoms propagating from left to right. Within 2 $\%$ accuracy, the final population in the target state $|1\rangle$ and the transmission are perfect. Except for the second Raman process, the velocity stays roughly constant signaling adiabatic evolution. During the second Raman process, as photons decay, the system cannot follow the dark state $\psi_d(x)$ adiabatically. Consequently, since the decay is not an adiabatic process, the velocity changes within this region. However, once leaving the cavity interaction zone, the initial velocity $v_0$ is restored, as will be discussed further at the end of this section. 

In the following two plots, (c) and (d), we display the same but for a condensate in the internal $|1\rangle$-state propagating in the negative direction. This time, since the cavity is empty the condensate traverses the first coupling pulses unaffected and reflect upon the Stark-shift pulse $W(x)$. Finally in (e) and (f), the results for negatively propagating condensates initially in the $|3\rangle$-state are shown. Due to the large detuning, $\Delta\approx\hbar5\Omega_0$, approximately no population transfer between the atomic states $|2\rangle$ and $|3\rangle$ takes place. Instead, the $G_p(x)$ pulse acts as a Stark-sfift and the condensate gets reflected by it. For the examples of (c)-(f), the transmitivity $T$ is of the order $10^{-6}$. If the atomic velocity is increased, atoms will at some point traverse the Stark-shift pulses. For atoms propagating from right to left, we have numerically seen that for velocities $v_0<-50$ m/s the atoms in the $|1\rangle$ state begin to pass the reflective $W(x)$ potential. At this velocity, the transmission is about 2 $\%$. For the employed system parameters, we do not find any tunneling resonances~\cite{mazer} within the velocity interval 5-50 cm/s. Beyond velocities $v_0<50$ cm/s, the $|1\rangle$-state atoms have enough kinetic energy to pass the central barrier. The reflective pulse $G_p(x)$ for $|3\rangle$ atoms is not as strongly reflective as $W(x)$ is for the $|1\rangle$ atoms, and consequently the critical velocity is instead around 35 cm/s. Again, both these velocities could be increased by using stronger lasers.   

\begin{figure}
\begin{center}
\includegraphics[width=8cm]{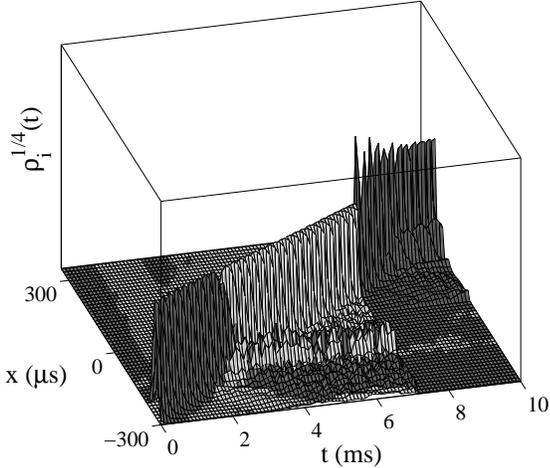}
\caption{Forth-root of the condensate propagation of the reduced densities $\sqrt[4]{\rho_1(t)}$ (dark gray surface) and $\sqrt[4]{\rho_3(t)}$ (light gray surface). The non-adiabatic effects, seen as population escaping the main wave packet, become clear in this plot by presenting the fourth-root of the densities. The parameters and conditions (given in the text) are the same as for fig.~\ref{fig2} (a) and (b).}
\label{fig3}
\end{center}
\end{figure}

The forth-root of the reduced densities, i.e. $\sqrt[4]{\rho_1(t)}$ and $\sqrt[4]{\rho_3(t)}$, for the positively propagating case are displayed in fig.~\ref{fig3} (the figure only gives one out of the 500 quantum Monte-Carlo trajectories). By showing the fourth-root of the densities, the small non-adiabatic contributions during the population transfers are seen. These appear as a small amount of population being trapped in the first STIRAP interaction region. During the second STIRAP, information leaks the system due to coupling of the cavity to its environment. The resulting decoherence acts as a reduction of the system wave function. This decoherence-induced collapse actually increases the efficiency of the diode. From fig.~\ref{fig3}, it is especially clear that the non-adiabatic contributions established during the first STIRAP process are suppressed during the wave function reduction. Due to the non-adiabaticity, the motional degrees-of-freedom of the condensate become entangled with the cavity field. However, once the field decays, the entanglement is lost and the motional wave packet of the condensate collapses at the same instant. This effect becomes more transparent at high velocities. For a condensate with an initial velocity of 50 $cm/s$, 74 $\%$ of the population remains in the dark state after the first STIRAP, but during the decoherence process of the second STIRAP the system relaxes almost perfectly back to its target state $|1\rangle$. Thus, even for  non-adiabatic evolution, the scheme works extremely well.      

It is known that nonlinearity can greatly suppress adiabaticity in the STIRAP scheme~\cite{nlinstirap}, as well as in more general settings~\cite{nlinlz}. For sufficiently strong non-linearity, e.g. increasing the atomic density or the scattering length $a_s$, the adiabatic potentials/energies form a loop-structure which prevents full adiabaticity regardless of how long the characteristic time-scale is~\cite{nlinstirap}. The breakdown of adiabaticity originating from atom-atom interaction was recently demonstrated in a double-well BEC system~\cite{bloch}. We have numerically checked the diode efficiency by varying the atom number $N$ (equivalent to keeping the density fixed while tuning the scattering length $a_s$). For the given value of $a_s$ and within the range between 10 000 to 2 000 000 atoms, we do not find substantial changes in the efficiency. Thus, we conclude that within this range, and for the present parameters, there are no loop-structures formed.

\section{Conclusion}
We have generalized the atom diode of ref.~\cite{diode} to apply for a condensate of ultacold atoms. The idea of~\cite{diode} relies on irreversible spontaneous emission which in a condensate will cause heating as well as decoherence and thereby breakdown of condensation. The present scheme reckon upon coherent coupling of the condensate with a cavity mode, which in return couples irreversibly to a photon reservoir. Decoherence in the condensate is then induced via the cavity field decay and should consequently be considerably suppressed. More precisely, we demonstrated the method on a mean-field level of the BEC, which does not fully take undesired quantum noise of the true atomic many-body system into account. Nevertheless, we believe that in terms of atomic condensates the present method should be superior the one of ref.~\cite{diode}, and photon reservoir noise has indeed been considered in the present model.

Utilizing experimentally relevant parameters, it was particularly shown how almost perfect efficiency could be reached. The system also gave an interesting example of non-locality and how the reservoir-induced wave packet reduction improves the diode performance, e.g. non-adiabatic contributions are annihilated by the collapse. The model Hamiltonian is, of course, an idealized case, and we have not taken; atomic many-body effects, heating due to thermal atoms, nor system fluctuations into account. On the other hand, the process time-scale is at the ms level, which is well below typical survival times of atomic condensates. It should be pointed out that the strength of the cavity decay $\kappa$ sets some conditions on repeatability of the diode; namely if a second condensate passes through the diode it is assumed that the cavity field has had enough time to relax to vacuum. With the $\kappa$ used in our examples this would certainly be true as the cavity field amplitude is approximately zero as soon as the condensate has left the interaction region. Finally, the present scheme relies on the STIRAP process, but we note that possible other adiabatic processes could be employed, for example a Landau-Zener transition~\cite{lz} between the two atomic states $|1\rangle$ and $|3\rangle$. However, such chirp-scheme would typically require changing of the internal atomic energies, which is most likely more difficult to accomplish experimentally. Therefor, the present STIRAP proposal is probably more relevant for current experimental setups.

\acknowledgments
The author acknowledges financial support from the Swedish research council--Vetenskapsr\aa det/VR.

\end{document}